\documentclass[conference]{IEEEtran}
\IEEEoverridecommandlockouts
\pdfoutput=1 % pdflatex hint for arxiv.org

\usepackage{amsmath,amssymb,amsfonts}
\usepackage{algorithmic}
\usepackage{graphicx}
\usepackage{textcomp}
\usepackage[cmyk, table]{xcolor}

\usepackage{ifpdf}

\ifpdf
\pdfoutput=1 % we are running pdflatex 
\pdftrue
\fi

\ifpdf
\else
  \usepackage[active]{srcltx}
\fi
\usepackage[utf8]{inputenc} % = latin1, latin9 (with euro sign), utf8
\usepackage[T1]{fontenc}
\usepackage{ae}
\usepackage{aecompl}
\usepackage{aeguill}

\usepackage{textcomp}

\usepackage{url}
\ifpdf

\else

\fi

\usepackage{xspace}
\usepackage{relsize}
\usepackage[protrusion=true,expansion=true,final]{microtype}
\usepackage{orcidlink}

\usepackage[nolist]{acronym}
\usepackage{tabto}
\usepackage{enumerate}
\usepackage{tikz}
\usetikzlibrary{arrows,arrows.meta}
\usepackage{pgfplots}
\pgfplotsset{compat=1.18}

\usepackage{comment}
\usepackage{tabularx}
\usepackage{array}
\usepackage{booktabs}
\usepackage{threeparttable}
\usepackage{wasysym}
\usepackage{ragged2e}
\usepackage{enumitem}
\usepackage{lscape}
\usepackage{pdflscape}
\usepackage{afterpage}

\usepackage{xurl}
\usepackage{hyperref}
\hypersetup{hidelinks,unicode}

\usepackage[style=ieee, backend=biber, url=true, hyperref, maxnames=5, minnames=1, maxcitenames=2, mincitenames=1,defernumbers=true]{biblatex}
\makeatletter
\let\blx@rerun@biber\relax
\makeatother

\usepackage{cleveref}

\DeclareBibliographyDriver{online}{%
  \usebibmacro{bibindex}%
  \usebibmacro{begentry}%
  \usebibmacro{author/editor+others/translator+others}
  \setunit{\labelnamepunct}\newblock
  \usebibmacro{title}%
  \newunit
  \printlist{language}%
  \newunit\newblock
  \usebibmacro{byauthor}%
  \newunit\newblock
  \usebibmacro{byeditor+others}%
  \newunit\newblock
  \printfield{version}%
  \newunit
  \printfield{note}%
  \newunit\newblock
  \printlist{organization}
  \newunit\newblock
  \usebibmacro{date}%
  \newunit\newblock
  \iftoggle{bbx:eprint}
    {\usebibmacro{eprint}}
    {}%
  \newunit\newblock
  \usebibmacro{url+urldate}%
  \newunit\newblock
  \usebibmacro{addendum+pubstate}%
  \setunit{\bibpagerefpunct}\newblock
  \usebibmacro{pageref}%
  \usebibmacro{finentry}}

\makeatletter
\renewcommand{\fnum@figure}{Figure~\thefigure}
\makeatother

\usepackage{xpatch}
\xpatchcmd\IEEEkeywords{---}{-}{}{}

\usepackage{ccicons}
\makeatletter
\def\ps@IEEEtitlepagestyle{
  \def\@oddfoot{\mycopyrightnotice}
  \def\@evenfoot{}
}
\def\mycopyrightnotice{
  {\footnotesize
    \begin{minipage}{0.8\textwidth}
    \centering
    Please cite as: \fullcite{selfref}.
    \end{minipage}
  }
}
\makeatother
\usepackage[shortcuts]{extdash} % Use \-/ for a breakable dash that does not prevent the remainer of the word to be hyphenated

\DeclareBibliographyCategory{selfref}
\addtocategory{selfref}{selfref}
\begin{acronym}
\acro{babe}[BABE]{Blind Assignment of Blockchain Extension}
\acro{fscn}[FSCN]{Food Supply Chain Network}
\acro{grandpa}[GRANDPA]{GHOST-based Recursive Ancestor Deriving Prefix Agreement}
\acro{gui}[GUI]{Graphical User Interface}
\acro{ide}[IDE]{Integrated Development Environment}
\acro{ocw}[OCW]{Off-Chain Worker}
\acro{sc}[SC]{Supply Chain}
\acro{scm}[SCM]{Supply Chain Management}
\acro{scn}[SCN]{Supply Chain Network}
\acro{hrmp}[HRMP]{Horizontal Relay-routed Message Passing}
\acro{xcm}[XCM]{Cross-Chain Messaging}
\acro{xcmp}[XCMP]{Cross-Chain Message Passing}
\acro{rbac}[RBAC]{Role Based Access Control}
\end{acronym}

\begin{document}

% Diferring from IEEE, IARIA requires 14 point Times New Roman in Bold (for the title)
% (like compsoc, but not with arabic numbering numbering that goes with compsoc)
\title{%
\bfseries\Large %
FoodFresh: Multi-Chain Design for an Inter-Institutional\\Food Supply Chain Network%
}

\author{
\IEEEauthorblockN{~\\[-0.4ex]\large %
Philipp Stangl and Christoph P.\ Neumann \orcidlink{0000-0002-5936-631X}%
\smallskip%
\normalsize}
\IEEEauthorblockA{%
Department of Electrical Engineering, Media and Computer Science\\
Ostbayerische Technische Hochschule Amberg-Weiden\\
Amberg, Germany\\
% Diferring from IEEE ("Email"), IARIA requires "e-mail":
e-mail: {\tt$\lbrace$p.stangl1\,|\,c.neumann$\rbrace$@oth-aw.de}
}
}

\maketitle

\begin{abstract}
We consider the problem of supply chain data visibility in a blockchain-enabled supply chain network.
Existing methods typically record transactions happening in a supply chain on a single blockchain and are limited in their ability to deal with different levels of data visibility.
To address this limitation, we present FoodFresh -- a multi-chain consortium where organizations store immutable data on their blockchains.
A decentralized hub coordinates the cross-chain exchange of digital assets among the heterogeneous blockchains.
Mechanisms for enabling blockchain interoperability help to preserve the benefits of independent sovereign blockchains while allowing for data sharing across blockchain boundaries.
\end{abstract}

% A list of IEEE Computer Society appoved keywords can be obtained at
% http://www.computer.org/mc/keywords/keywords.htm

\section{Introduction} \label{s:introduction}
The food industry comprises companies dedicated to manufacturing and processing raw materials and semi-finished products from agriculture, forestry, and fishing. In recent years, food supply chains have progressed from shorter, independent to more unified, coherent relationships among supply chain participants \cite{bourlakis2008food}. Developing long-term, and collaborative relationships requires evolutionary technological solutions to simultaneously retain a competitive edge. 

Blockchain technology is considered a way to increase supply chain visibility, support fraud detection and provide supply chain optimization. Current applications of blockchain technology in food supply chain management, e.g., IBM Food Trust \cite{ibmFoodTrust2019}, rely mainly on a single distributed ledger. The implications on supply chain networks are twofold: (i) organizations participating in multiple supply chains must share their data on multiple blockchains, and (ii) participants may see information originally not intended for them because all participants can view every transaction on a distributed ledger. In a single-chain approach with just one ledger, all data would be shared publicly will all other chain participants.

The motivation and novelty of the multi-chain is required by achieving controlled transparency, i.e., to enable all parties to control visibility of data based on two levels of chains. Each participant is provided with a chain of type \emph{permissioned}, and sharing data is provided by an additional chain of type \emph{public}. The permissioned chains are subject to a \ac{rbac} mechanism, thus, its information is hidden from the public and accessible to all users that belong to an organization. Providing organizations each with their own permissioned chain, interconnecting them as a federated ecosystem with a public chain also simplifies the addition or removal of individual organizations from the overall ecosystem with minimal impact.

In this paper, we propose FoodFresh -- a multi-chain approach for inter-institutional supply chain networks, allowing organizations to store immutable data on their blockchain. A decentralized hub coordinates the cross-chain communication among the heterogeneous blockchains.  The hub further ensures that all parties comply with the overarching rules of the consortium. 

The remainder of the paper is organized as follows: in Section~\ref{s:related_work}, a selection of related work is presented. Subsequently, an overview of the relevant technology is given in Section~\ref{s:background}. Next, Section~\ref{s:chainFresh} discusses our proposal with the design rationale. We conclude the paper in Section~\ref{s:conclusion}, followed by the references at the end. 

\section{Related work} \label{s:related_work}
Recently, various solutions for blockchain-enabled supply chains have been proposed. For instance, \citeauthor{longo2019blockchain} have presented a software connector to connect an Ethereum-like public blockchain with an enterprise information system \cite{longo2019blockchain}. The software connector allows companies to share information with their partners with different levels of visibility. \citeauthor{schulz2018multichain} \cite{schulz2018multichain} have proposed a blockchain-enabled distributed supply chain. Their main idea is a network-centric design, which incorporates domain-specific blockchains for handling specific business processes and a hub or main blockchain that connects the blockchains to communicate with each other. 

Polkadot uses a hybrid consensus model, separating block production (\ac{babe}) from finality (\ac{grandpa}). This allows for blocks to be rapidly produced and finalized at a slower pace without risking slower transaction speeds or stalling.  Polkadot provides cross-chain communication with arbitrary data. Parachains communicate through the \ac{xcmp} protocol, a queuing communication mechanism based on a Merkle tree. \ac{xcmp} is designed to communicate arbitrary messages between parachains. Messages are sent together with the next parachain block (short: parablock), while the relay chain blocks include only the proof of postage. All messages must be processed in proper order, for which a chain of Merkle proofs is used. However, \ac{xcmp} is still under development. Therefore, the stop-gap protocol is \ac{hrmp}. As soon as \ac{xcmp} is fully developed, it can replace \ac{hrmp}. The primary difference between the two is the data stored on the relay chain. In \ac{hrmp}, the relay chain stores the full message with its payload. \ac{xcmp}, on the other hand, will only store a reference to the payload. The target parachain will be responsible for decoding the message payload.

From the perspective of inter-institutional supply chains, FoodFresh extends our previous work on inter-institutional cooperation 
\cite{Neumann2013dissBook, NRDL09deus, NeLe12alphaFlow, NeLe09dmps} that was focused on healthcare, in which central organizations from primary care and secondary care act as leaders and hubs of cooperation.
The FoodFresh scenario extends our perspective to more decentralized and autonomous institutional cooperation in a food supply chain network, without central protagonists.

\section{Background} \label{s:background}
This section provides a brief overview of the different relevant technologies: Section~\ref{s:fscn} describes the characteristics of food supply chain networks, Section~\ref{s:blockchain_types} presents different types of blockchain technology, and Section~\ref{s:blockchain_interoperability} different blockchain interoperability approaches.

\subsection{Food Supply Chain Network} \label{s:fscn}
A supply chain is an interconnection of organizations, activities, resources, people, and information. Organizations along a food supply chain are dedicated to growing and processing raw materials (e.g., fruits) and semi-finished products (e.g., fruit juices) for delivery to the end customer. Food supply chains are complex and affected by various factors, such as the sociopolitical environment \cite{van2005innovations}. Regulatory bodies, such as the US Department of Agriculture (USDA), aim to protect consumer health and increase economic viability. Thus, they release frequent updates to ensure their criteria are met by food supply chains.

In a \ac{fscn}, more than one supply chain and more than one business process can be identified, both parallel and sequential in time. The parties involved in the business processes depend on the type of \ac{fscn}. This article considers a \ac{fscn} for fresh agricultural products.  

\begin{figure}[ht]
    \centering
    \includegraphics[page=1,scale=0.23]{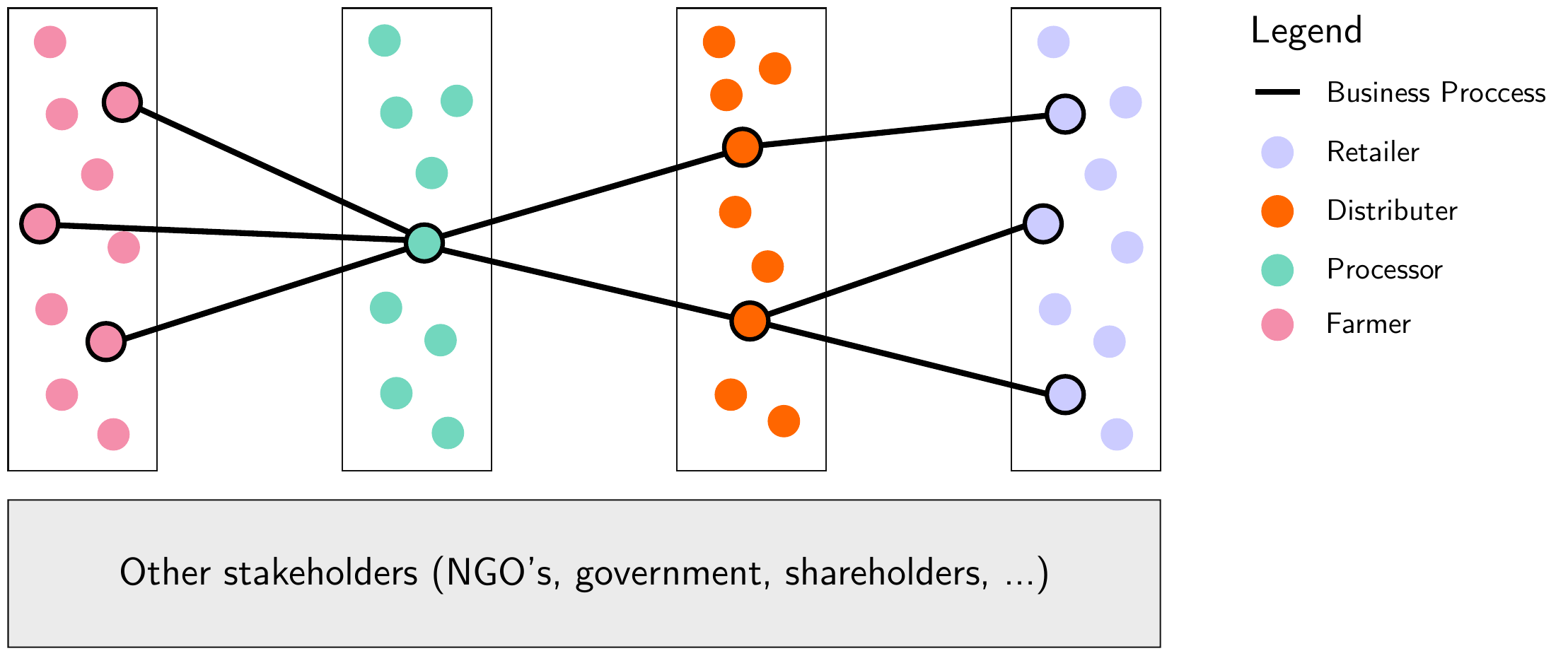}
	\caption{Schematic diagram of an \ac{fscn} (based on \citeauthor{van2005innovations}~\cite{van2005innovations})}
	\label{fig:food_supply_network}
\end{figure}

\citeauthor{van2005innovations} have identified farmers, retailers, and their logistics service suppliers as parties involved in a \ac{fscn} for fresh agricultural products \cite{van2005innovations}. Figure~\ref{fig:food_supply_network} depicts such a supply chain at the organization level within the context of a \ac{fscn} for fresh agricultural products. Each organization is positioned in a product lifecycle stage and belongs to at least one supply chain. That means an organization can have multiple suppliers and customers at the same time and over time. Figure~\ref{fig:food_supply_network} visualizes this by showing the perspective of the processor (bold lines), who has multiple connections to distributors and farmers. Other stakeholders, such as nongovernmental organizations, governments, and shareholders, are indirectly involved at each stage of the product lifecycle.

\subsection{Blockchain Types} \label{s:blockchain_types}
There are three different types of blockchain systems \cite{zheng2017overview}. Public blockchains are considered permissionless because, in principle, everyone can attend the consensus process and read the stored data. The application of public blockchains has several use cases, including cryptocurrencies and document validation. In a consortium blockchain, an elected group of participants is allowed to attend the consensus process. The stored data may be read by selected members or by the public. Supply chain and research environments are two exemplary use cases for this type of blockchain. In a private blockchain, all participants belong to the same organization, and the public cannot access the system. Two use cases for this final blockchain type are banking and asset ownership. Private and consortium blockchains are considered permissioned blockchains because, in both cases, only a limited group can attend the consensus process.

\subsection{Blockchain Interoperability} \label{s:blockchain_interoperability}
Blockchain interoperability involves the ability of independent distributed ledger networks to communicate with each other. Various approaches have been established to provide blockchain interoperability, resulting in a highly fragmented market \cite{belchior2021survey}.
\citeauthor{belchior2021survey} were the first to conduct a systematic literature review in \cite{belchior2021survey} on blockchain interoperability solutions:
Their resulting Blockchain Interoperability Framework categorizes interoperability solutions into three categories: 1) interoperability across public blockchains (public connectors),  2) independent blockchains that interoperate among each other (blockchains of blockchains), and finally, 3) approaches that neither fit into the public connectors nor blockchains of blockchains category (hybrid connector). 

\section{FoodFresh} \label{s:chainFresh}
In this section, we describe a consortium blockchain for a food supply chain network for interoperability and controlled transparency. Section~\ref{s:chainfresh_approach}, introduces the approach. The three tiers of the system architecture are described in the following sections: the presentation tier in Section~\ref{s:presentation_tier},  the application tier in Section~\ref{s:application_tier}, and the relay tier in Section~\ref{s:relay_tier}. In Section~\ref{s:substrate_framework}, we offer a concise introduction to the Substrate Framework. The subsequent Section~\ref{s:deployment}, delineates the details concerning the deployment process. Finally, Section~\ref{s:limitations} addresses the limitations of our proposed approach.

\subsection{FoodFresh Approach} \label{s:chainfresh_approach}
The FoodFresh approach provides an implementation of the multi-chain approach (Section~\ref{s:related_work}). The blockchain consortium comprises a multi-chain ecosystem for organizations. Each organization is allowed to participate in the consensus process. 
A permanent and shared record of food system data connects participants across the food supply chain network. This is done through the use of a main blockchain, called relay chain. The sole purpose of the relay chain is to coordinate and share appropriate data and ensure all parties are complying with overarching rules.
Each organization can set up and manage its own permissioned blockchain, which keeps full control over the data to itself.
Within a single permissioned blockchain for an organization, data is shared between different users that belong to that organization, but not to inter-institutional parties.
Via the public relay, the FoodFresh approach allows them to share immutable and accurate data with other participants in the inter-institutional supply network.
This also allows for the addition or removal of individual organizations from the overall ecosystem with minimal impact.

\subsection{System Architecture} \label{s:system_architecture}
FoodFresh, as a distributed system, is a composition of three tiers. 
This section will outline each of the three tiers. The presentation tier in Section~\ref{s:presentation_tier}, the application tier in Section~\ref{s:application_tier}, and finally the relay tier in Section~\ref{s:relay_tier}. Figure~\ref{fig:system_architecture} depicts the system architecture for two interoperating supply chain organizations. 

\subsubsection{Presentation Tier} \label{s:presentation_tier}
To provide the user with convenient access to the FoodFresh system, the presentation tier is responsible for interacting with the application tier through a websocket connection. 
Any websocket-capable client or device can communicate with the endpoints exposed by the application tier. The user interacts with a \ac{gui} to manage the permissions of participating members, register shipments and products, and trace shipments along the supply chain. A browser extension is required to manage blockchain accounts and to sign transactions within those accounts.

\subsubsection{Application Tier} \label{s:application_tier}
\begin{figure}[h!t]
\centerline{\includegraphics[scale=0.6]{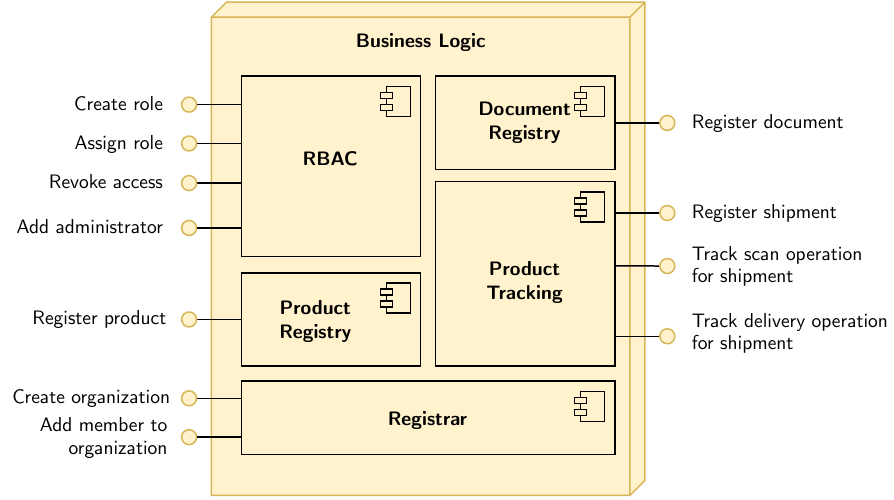}}
\caption{Overview of the business logic, decomposed into five pallets}
\label{fig:business_logic}
\end{figure}

The application tier encompasses application-specific blockchains (the parachains) that allow organizations to join with their blockchain, where they can store immutable data. Through this, organizations can create products and shipments. A shipment's storage and transportation conditions can be monitored and tracked through the supply chain. The business logic is decomposed in tightly coupled modules called pallets. Figure~\ref{fig:business_logic} depicts the business logic pallets, each with its provided functionality that can be invoked via transactions on the parachain. Additionally, an \ac{ocw} is used to communicate the latest shipment status with the external world. With the subsystem Cumulus, parachains can send and receive cross-chain messages and enable validators to validate their state transitions.
\ac{rbac}, formalized by \citeauthor{ferraiolo2003role} \cite{ferraiolo2003role}, has become the predominant model for user access control. \ac{rbac} is used in the FoodFresh approach to control the access in terms of who can submit transactions. The \textit{rbac} pallet maintains an on-chain registry of roles and the users to which those roles are assigned. A role is a tuple with the name of a pallet and a permission that qualifies the level of access granted by the role. A permission is an enumeration with the variants \textit{Execute} and \textit{Manage}. The \textit{Execute} permission allows a user to invoke a pallet’s dispatchable functions. The \textit{Manage} permission allows a user to assign and revoke roles for a pallet, and also implies the \textit{Execute} permission. Access control validation is done within the transaction pool of a parachain.

\afterpage{%
\begin{landscape}
\begin{figure}[!ht]
  \centering
  \includegraphics[height=0.95\textheight]{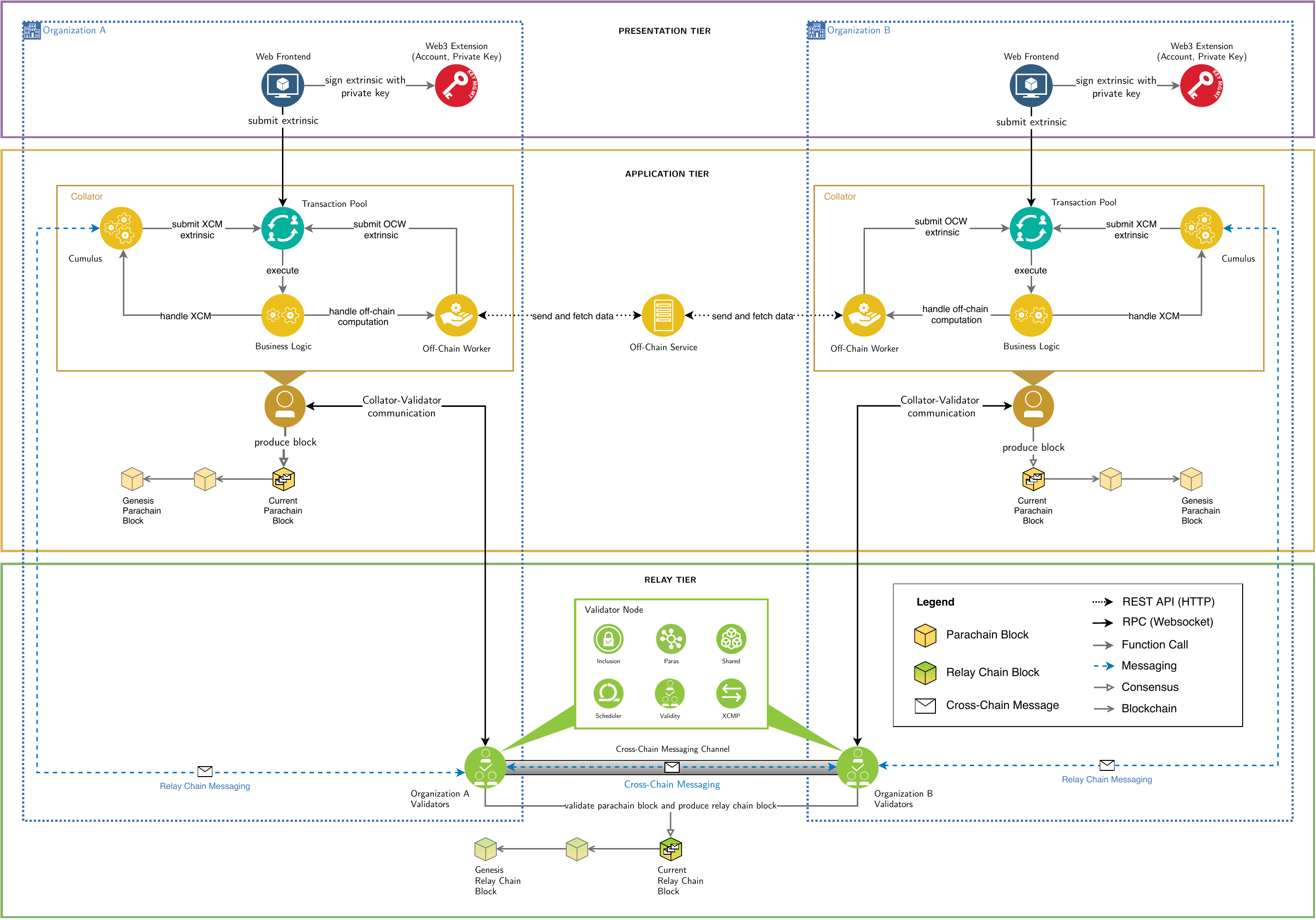}
  \caption{Overview of our approach. The architecture is composed of three tiers: presentation, application, and relay. }
	\label{fig:system_architecture}
\end{figure}
\end{landscape}%
}

\subsubsection{Relay Tier} \label{s:relay_tier}
The relay chain, in the relay tier, is the essential hub in the network of heterogeneous blockchains, the parachains. The relay chain provides parachains with parablock validation and allows them to communicate with each other using the \ac{xcm} format for cross-chain messaging.

Validators are the actors of the relay chain and have three responsibilities: (1) to verify that the information contained in parablocks is valid, such as the identities of the transacting parties, (2) to participate in the consensus mechanism to produce the relay chain blocks based on validity statements from other validators, and (3) to handle cross-chain messages. For validators to fulfill their responsibilities, they are equipped with six primary runtime modules. The \textit{inclusion} module handles the inclusion and availability of parablocks. In addition, \textit{shared} manages the shared storage and configurations for other validator modules. The \textit{paras} module manages the chain-head and validation code for parachains. The \textit{scheduler} is responsible for parachain scheduling, as well as validator assignments for the consensus mechanism. The \textit{validity} module addresses secondary checks and disputes resolution for available parablocks. Finally, the \textit{\ac{xcmp}} module handles cross-chain messages and ensures that the messages are relayed to the receiving parachain. 

An integral part of cross-chain communication is the establishment of a cross-chain messaging channel between the validators of two communicating parachains. \citeauthor{burdges2020overview} \cite{burdges2020overview} have stated that a messaging channel aims to guarantee four things: ``First that messages arrive quickly; second that messages from one parachain arrive to another in order; third that arriving messages were indeed sent in the finalized history of the sending chain; and fourth that recipients will receive messages fairly across senders, helping guarantee that senders never wait indefinitely for their messages to be seen''.

The act of removing an organization from the ecosystem does not necessitate the elimination of its associated parachain. This concept is facilitated by the existence of a systematic protocol, specifically the modification of the relay chain validator registry. Through the processes of registration and deregistration, organizations are added to and removed from this registry. Structurally, this registry is characterized as a hash map, a data structure that comprises paired elements: a unique identifier (ID) and the corresponding parachain ID. It should be noted that the relationship between organizations and their parachains is fundamentally non-destructive, meaning that the alterations in the organization's status within the ecosystem do not directly impinge on the existence of the related parachain.

%Development
\subsection{Substrate Framework} \label{s:substrate_framework}
FoodFresh is built with substrate \cite{substrateio}, a modular framework for building blockchains. A nontechnical reason for using substrate is its flexibility. Organizations must be able to adapt their blockchain system to meet the supply chain compliance requirements of regulatory bodies. Regulations happen frequently, especially in food supply chains, as shown in Section~\ref{s:fscn}. Due to the modular nature of substrate-based blockchains, developers have the necessary freedom to swap or add modules to their blockchain runtime.

Technical reasons include the chosen programming language, the software design, and the off-chain abilities. Substrate is implemented in the programming language Rust, which aims to provide performance (comparable to C++), reliability, and better means of productivity. In terms of reliability, Rust manages resources (including memory, files, network, and thread) and avoids problems, such as resource leaks or data races. Finally, for productivity, Rust provides \ac{ide} support and type inspections. Furthermore, substrate is generic by design, meaning transactions are abstracted to so-called extrinsics (things that happen outside the chain) and intrinsics (things that happen inside the chain). Transactions are stored as binary large objects. As a result, users can transfer and store any type of data on the blockchain. 

Nonetheless, with FoodFresh as a permissioned blockchain, concerns about off-chain processes need to be raised. For instance, \citeauthor{HELLIAR2020102136} have made the assumption that ``off-chain processes may become a major barrier for permissioned blockchains'' \cite{HELLIAR2020102136}. Using substrate, off-chain data can be queried or processed before it is included in the on-chain state through \ac{ocw}, a collator node subsystem that allows for the execution of long-running and possibly nondeterministic tasks. Moreover, an \ac{ocw} does not influence the block production time.

\subsection{Deployment} \label{s:deployment}
FoodFresh requires validator nodes for the relay chain and collator nodes for the parachains to be set up by the organizations participating in a supply chain network. Nodes can be deployed locally or remotely via a cloud service provider, such as Amazon Web Services. 
Before parachains can participate in cross-chain communication, they need to be registered on the relay chain. The following rule is defined in the Collator Protocol \cite{collatorProtocol}, which implements the network protocol for the Collator-to-Validator networking: To accept $n$ parachain connections, $n + 1$ validator nodes need to run on the relay chain. For the FoodFresh prototype, two relay chain nodes are started to connect one parachain node. Further, the relay chain needs to obtain the hex-encoded parachain's genesis state (exported from a collator node) and the WebAssembly runtime validation function to validate parablocks.

\subsection{Limitations} \label{s:limitations}
While the FoodFresh approach offers a comprehensive framework for leveraging blockchain technology in food supply chain management, there are several potential limitations and areas of concern, including:\\
\textit{Scalability}: Parachains might face scalability challenges, depending on the scale of the organizations involved and the number of transactions. These issues are typically dependent on their specific implementation, the consensus mechanism used, and the volume of transactions they handle. If an organization's parachain is not optimized to handle large quantities of data or high transaction throughput, it could become a bottleneck that slows down the overall system's performance.\\
\textit{Complexity of Implementation}: The FoodFresh approach, with each organization having its own blockchain and one relay chain for cross-communication, increases the complexity of the system compared to commonly used single blockchains. This presents significant challenges in terms of maintenance and understanding the system for non-technical stakeholders.\\
\textit{Adoption Challenges}: Organizations might be reluctant to adopt the proposed approach due to perceived risks, lack of understanding, or the costs involved in implementation and training. 
\\%
\textit{Evaluation}: The software prototype is implemented and available on GitHub\footnote{\href{https://github.com/cyberlytics/FoodFresh}{https://github.com/cyberlytics/FoodFresh}} under Apache License 2.0.
A key part of designing a supply-chain network is ensuring the network is versatile enough to cope with future risks.
The current solutions to analyze and mitigate endogenous risks lack continuous monitoring, as a result, risks from irregularities (e.g.\ abnormal order quantities by retailers) remain mostly undetected.
Thus, a core part of our future evaluation is to answer whether we can develop an approach to detect abnormal activity in a multi-chain scenario.
Our plans will focus on capturing the variability of transfer volume in cross-chain messaging in order to detect abnormal activity in blockchain-enabled inter-institutional supply chain networks.

\section{Conclusion and Future Work} \label{s:conclusion}
Developing long-term and increasingly collaborative relationships among supply chain participants requires advanced technological solutions to retain a competitive edge. Blockchain is presented as a promising technology that might increase supply chain visibility and improve efficiency.  We have presented FoodFresh -- a multi-chain consortium for an inter-institutional food supply chain network. 
This approach overcomes the challenges associated with current approaches (e.g., IBM Food Trust), such as lack of controlled transparency and restricted interoperability among supply chain participants. By implementing a multi-chain consortium with an overseeing decentralized hub, FoodFresh allows organizations to maintain their independent blockchains, thereby preserving data sovereignty and enabling effective data exchange across blockchain boundaries. 
The design approach used for FoodFresh could apply to other networks that require the distribution or transfer of sensitive data. Future work could apply the approach to other industries, for instance, healthcare. The safe and secure transfer of patient health records or other sensitive information between healthcare providers, insurance companies, and the patients themselves is a major concern in the healthcare industry. The presented approach could allow each party to maintain control over their data while enabling necessary data sharing.

\vspace{1cm}

% ======== References =========
\begingroup
%\raggedright
\sloppy
\microtypesetup{protrusion=false}
%\vspace{\fill} % References auf nächste Spalte bringen
%\renewcommand*{\bibfont}{\small}
%\setlength{\labelnumberwidth}{0.45cm}
\printbibliography[notcategory=selfref]
%\vspace{\fill} % Nötig, weil sonst zwischen References und dem ersten Eintrag ein hässlicher Whitespace entsteht
%\vspace{12pt}
\endgroup 

\end{document}